\documentclass{article}
\usepackage{spconf,amsmath,graphicx,hyperref}
\usepackage{amssymb}
\title{DriftAudio: Marginal Drifting for Distributional Post-Training of One-Step Text-to-Audio Generators}
\name{Xingyu Chen, Fei Ma, Sipei Zhao}
\address{Centre for Audio, Acoustics and Vibration, Faculty of Engineering and IT\\
University of Technology Sydney, Ultimo, NSW 2007, Australia}
\begin{document}
\ninept
\maketitle

\begin{abstract}
Recent one-step text-to-audio (TTA) models substantially reduce inference cost, yet their generated distributions can still be improved through post-training.
We propose \textbf{DriftAudio}, a distributional post-training method that adapts Drifting to pretrained one-step TTA generators.
Applying Drifting condition-wise is challenging under free-form text conditioning, where only one or a few real samples are typically available for a particular condition.
DriftAudio instead performs Drifting on the marginal audio distribution while retaining text-conditioned generation.
The drifting field is estimated in a frozen audio feature space using resampled real samples, a rolling bank of generated samples, and the current batch of generated samples.
The resulting Drifting field provides a detached training target for updating only the generator, keeping the original one-step inference procedure.
On AudioCaps, starting from MeanAudio, DriftAudio reduces FAD and FD by
$33.9\%$ and $17.6\%$, respectively, while also improving KL and CLAP.
Starting from FdAudio, it further reduces FAD, FD, and KL, with trade-offs in IS and CLAP.
These results demonstrate the effectiveness of marginal distributional post-training for one-step TTA generation.
\end{abstract}
\begin{keywords}
text-to-audio generation, one-step generation, distributional post-training, drifting
\end{keywords}

\section{Introduction}
\label{sec:introduction}

Text-to-audio (TTA) generation aims to synthesize realistic audio aligned with free-form natural-language descriptions~\cite{
kreuk2023audiogen}.
Diffusion- and flow-based models have substantially improved generation quality~\cite{
liu2023audioldm,
liu2024audioldm2,
hung2026tangoflux,
guan2024lafma}, but high-quality inference commonly relies on iterative sampling with repeated neural network evaluations (NFEs).
This motivates low-NFE TTA generation, with strict one-step generation requiring only a single NFE.

Recent low-NFE TTA methods reduce sampling cost by compressing iterative sampling into a small number of NFEs.
Consistency-based methods learn consistent mappings across states along diffusion trajectories, enabling one- or few-step synthesis~\cite{
bai2024consistencytta,
liu2024audiolcm,
saito2025soundctm}.
Trajectory rectification improves few-step generation by simplifying or straightening the sampling path~\cite{
zhao2025audioturbo,
liu2025flashaudio}.
More recently, MeanAudio~\cite{li2026meanaudio} adopts the MeanFlow objective~\cite{geng2025meanflow} to learn average flow dynamics over finite time intervals, enabling strict $1$-NFE generation.
Despite these advances, generators can still benefit from further quality refinement, motivating post-training refinement.

Distributional post-training provides a complementary route.
Rather than introducing a new inference-time sampler, it directly refines the distribution induced by a pretrained generator.
FdAudio~\cite{huang2026fdaudio}, for example, post-trains MeanAudio~\cite{li2026meanaudio} by matching real and generated audio distributions across multiple frozen audio feature spaces using Fr\'echet objectives~\cite{yang2026fdloss} based on first- and second-order moments.

Drifting~\cite{deng2026drifting} provides a sample-based alternative to moment-based distribution matching by constructing a
distribution-dependent field from real and generated samples through kernel-weighted interactions.
Drifting has recently been explored for one-step speech enhancement~\cite{xu2026driftse}, but its use for TTA remains unexplored.
A direct condition-wise application to TTA would require estimating a separate conditional data distribution for each text condition.
This is challenging because each free-form text condition is typically associated with only one or a few real audio samples.
Our key observation is that, after conditional generation has been learned, distributional refinement can be performed on the marginal output distribution without estimating a separate data distribution for each text condition.

Building on this observation, we propose \textbf{DriftAudio}, a distributional post-training method for pretrained one-step TTA generators.
Generation remains text-conditioned, while distribution matching is performed marginally in a feature space.
Real feature samples are resampled at each step from the full training set, while the generator-induced distribution is represented by a first-in, first-out (FIFO) bank initialized with pre-generated samples and refreshed using a small batch of current generated samples.
The drifting field is then estimated between the current generated samples and the real and generated reference samples, and converted into a detached target to update only the generator while preserving one-step inference.

Our contributions are threefold:
(i) We formulate Drifting as a marginal distributional post-training objective for one-step TTA generators, avoiding the need to estimate a separate real-audio distribution for each text condition;
(ii) we develop a practical estimator using resampled real features
and a FIFO bank of generated features;
(iii) we demonstrate on AudioCaps that DriftAudio improves FAD, FD, and KL from both starting checkpoints, while retaining competitive text–audio alignment.
To the best of our knowledge, this is the first application of Drifting to TTA generation.
Audio demos are available on the project page. \footnote{\url{https://xingyuaudio.github.io/driftaudio-demo/}}
Code and model weights will be released upon acceptance.

\begin{figure*}[t]
    \centering
    \includegraphics[width=\textwidth]{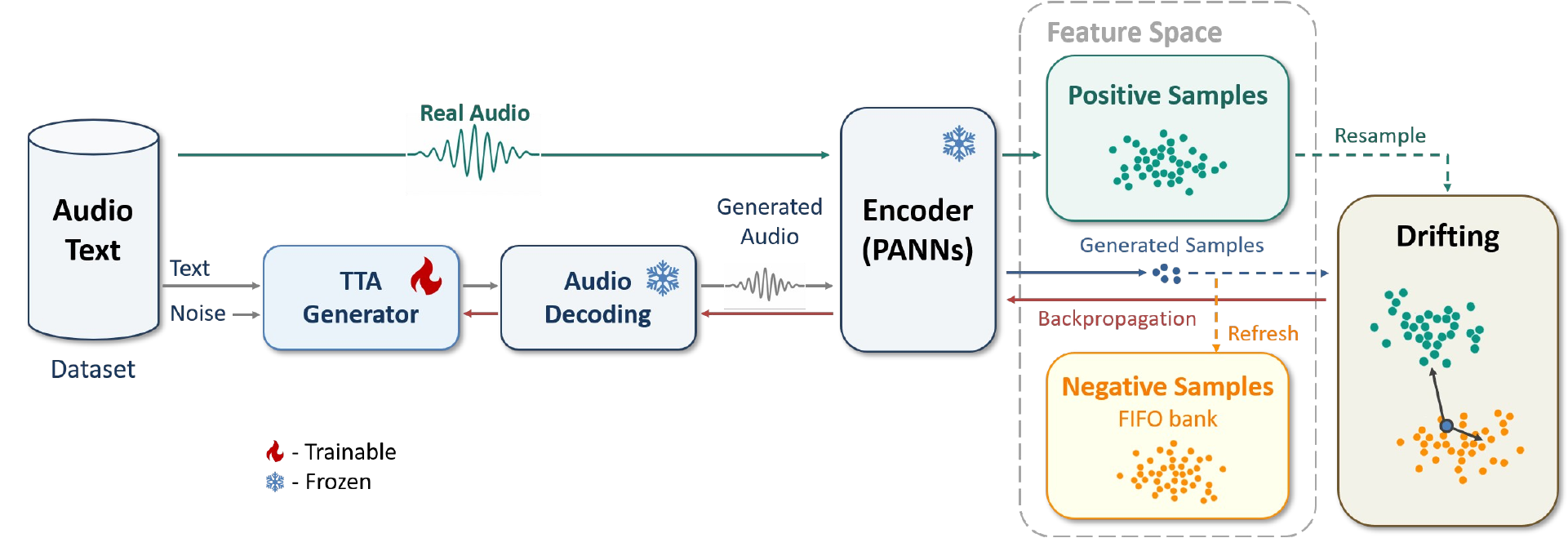}
    \caption{
Overview of DriftAudio.
The one-step generator remains text-conditioned, while Drifting is
performed on the marginal audio distribution in feature space.
At each optimization step, real features are resampled from the training set, while generated reference features are maintained in a FIFO bank initialized with pre-generated samples and updated using the current generated samples.
    }
    \label{fig:architecture}
\end{figure*}

\section{Preliminaries}
\label{sec:preliminaries}

\subsection{Pretrained One-Step TTA Generator}

We use the MeanAudio architecture~\cite{li2026meanaudio}, which enables
one-step TTA generation using the MeanFlow objective~\cite{geng2025meanflow}.
Unlike standard flow matching~\cite{lipman2022flow}, which models instantaneous velocity, 
MeanFlow models the average velocity over a finite interval $[r,t]$:
\begin{equation}
\mathbf u(\mathbf z_t,r,t,c)
=
\frac{1}{t-r}
\int_r^t
\mathbf v(\mathbf z_\tau,\tau,c)\,\mathrm d\tau,
\label{eq:average_velocity}
\end{equation}
where $\mathbf{z}_t$ denotes the latent state at time $t$, $c$ denotes the text condition and $0\leq r<t\leq 1$.
By definition, the displacement over this interval is
$(t-r)\mathbf u$, giving
$\mathbf z_r=\mathbf z_t-(t-r)\mathbf u(\mathbf z_t,r,t,c)$.
Taking the full interval from the noise endpoint $t=1$ to the data
endpoint $r=0$ yields a one-step latent generator:
\begin{equation}
\hat{\mathbf z}
=
G_\theta(\boldsymbol{\epsilon},c)
=
\boldsymbol{\epsilon}
-
\mathbf u_\theta(\boldsymbol{\epsilon},0,1,c),
\qquad
\boldsymbol{\epsilon}\sim\mathcal N(\mathbf 0,\mathbf I).
\label{eq:one_step_generation}
\end{equation}
The generated latent is converted to a waveform
$\hat{\mathbf a}=\mathcal D(\hat{\mathbf z})$, where $\mathcal D$
denotes the audio decoding pipeline (VAE~\cite{huang2023makeanaudio2} and vocoder~\cite{lee2022bigvgan}).

\subsection{Distribution Matching with Drifting}

Drifting~\cite{deng2026drifting} formulates generative learning by evolving the generator-induced distribution $q_\theta$ toward the data distribution $p_{\mathrm{data}}$ through a distribution-dependent field.
For a generated sample $\mathbf x\sim q_\theta$, the drifting field is
\begin{equation}
\begin{aligned}
\mathbf F_{p_{\mathrm{data}},q_\theta}(\mathbf x)
&=
\frac{
\mathbb E_{\mathbf y^{+}\sim p_{\mathrm{data}}}
\left[
k(\mathbf x,\mathbf y^{+})
(\mathbf y^{+}-\mathbf x)
\right]
}{
\mathbb E_{\mathbf y^{+}\sim p_{\mathrm{data}}}
\left[
k(\mathbf x,\mathbf y^{+})
\right]
}
\\[-1mm]
&\quad-
\frac{
\mathbb E_{\mathbf y^{-}\sim q_\theta}
\left[
k(\mathbf x,\mathbf y^{-})
(\mathbf y^{-}-\mathbf x)
\right]
}{
\mathbb E_{\mathbf y^{-}\sim q_\theta}
\left[
k(\mathbf x,\mathbf y^{-})
\right]
},
\end{aligned}
\label{eq:drifting_field}
\end{equation}
where $\mathbf y^{+}$ and $\mathbf y^{-}$ denote positive (real) and
negative (generated) reference samples drawn from
$p_{\mathrm{data}}$ and $q_\theta$, respectively, and
$k(\cdot,\cdot)$ denotes the similarity kernel.
For conditional generation, the construction can be applied to $p_{\mathrm{data}}(\cdot\mid c)$ and $q_\theta(\cdot\mid c)$.
In practice, the expectations are approximated empirically using finite samples, and the drifting field may be evaluated in a feature space.

\section{DriftAudio}
\label{sec:method}

DriftAudio post-trains a text-to-audio generator by applying Drifting~\cite{deng2026drifting} to its marginal audio distribution in a frozen feature space.
As illustrated in Fig.~\ref{fig:architecture}, each generated sample remains conditioned on its text input, while real and generated samples collected across text conditions are used to estimate the marginal drifting field.
A frozen audio encoder defines the feature space in which the field is constructed, and only the generator parameters are updated.

\subsection{Marginal Drifting under Free-Form Conditioning}
\label{sec:marginal_drifting}

For text-conditioned TTA, applying Drifting at the condition-specific level requires estimating $p_{\mathrm{data}}(\cdot\mid c)$.
Since $p_{\rm data}(\cdot|c)$ cannot be reliably estimated from one or a few samples per free-form text condition, we instead apply Drifting to the marginal distributions.
Given text conditions $\{c_n\}_{n=1}^{N}$, the marginal data and
generator-induced distributions are approximated by
\begin{equation}
\begin{aligned}
p_{\mathrm{data}}(\cdot)
&\approx
\frac{1}{N}
\sum_{n=1}^{N}
p_{\mathrm{data}}(\cdot\mid c_n),
&
q_\theta(\cdot)
&\approx
\frac{1}{N}
\sum_{n=1}^{N}
q_\theta(\cdot\mid c_n).
\end{aligned}
\label{eq:marginal_distribution}
\end{equation}
Hereafter, $p_{\mathrm{data}}$ and $q_\theta$ denote these marginal
audio distributions.

We approximate these marginal distributions with finite real and generated samples collected across text conditions, and estimate the corresponding drifting field in Eq.~\eqref{eq:drifting_field} in a frozen feature space defined by an audio encoder $\phi$.
In this feature space, the drifting field is constructed from three sets: positive samples, negative samples, and the current generated samples.
We define the positive and negative sets as
\begin{equation}
\mathcal Y^{+}
=
\left\{
\mathbf y_i^{+}=\phi(\mathbf a_i)
\right\}_{i=1}^{N_{+}},
\qquad
\mathcal Y^{-}
=
\left\{
\mathbf y_i^{-}=\phi(\hat{\mathbf a}_i)
\right\}_{i=1}^{N_{-}}.
\label{eq:reference_samples}
\end{equation}
At each optimization step, $\mathcal Y^{+}$ is resampled without replacement from the full set of real training features, while
$\mathcal Y^{-}$ is maintained as a FIFO bank initialized with pre-generated features and refreshed using the current generated
samples, which are defined as
\begin{equation}
\mathcal X
=
\left\{
\mathbf x_i=\phi(\hat{\mathbf a}_i)
\right\}_{i=1}^{Q}.
\label{eq:generated_samples}
\end{equation}
All feature samples lie in $\mathbb R^{D}$, where $D$ denotes the feature dimension.
Here, $\mathcal Y^{-}$ is maintained as a bank of detached features from recently generated samples, avoiding regeneration of a large set at every step, whereas $\mathcal X$ contains the current generated samples on which the field is evaluated.
The sets $\mathcal X$, $\mathcal Y^{+}$, and $\mathcal Y^{-}$ are then used to construct the marginal drifting field.

Following Drifting~\cite{deng2026drifting}, let $w_{ij}^{+}$ and
$w_{ij}^{-}$ denote the kernel-derived weights between $\mathbf x_i$
and the samples in $\mathcal Y^{+}$ and $\mathcal Y^{-}$,
respectively, with $S_i^{+}=\sum_j w_{ij}^{+}$ and
$S_i^{-}=\sum_j w_{ij}^{-}$.
The corresponding denominator-cleared marginal drifting field is
\begin{equation}
\mathbf F_i
=
S_i^{-}\sum_j w_{ij}^{+}(\mathbf y_j^{+}-\mathbf x_i)
-
S_i^{+}\sum_j w_{ij}^{-}(\mathbf y_j^{-}-\mathbf x_i).
\label{eq:marginal_drift}
\end{equation}

\subsection{Drifting-Based Post-Training}
\label{sec:drift_posttraining}

The previous subsection defines the marginal drifting field $\mathbf F_i$ in feature space.
We convert this field into a generator training objective by defining a detached target for each generated feature,
\begin{equation}
\mathbf t_i
=
\operatorname{sg}
\left(
\mathbf x_i+\mathbf F_i
\right),
\label{eq:drift_target}
\end{equation}
where $\operatorname{sg}(\cdot)$ denotes stop-gradient.

To account for the numerical scale of the feature space, we normalize the regression by the average distance between the current generated samples and the reference samples,
\begin{equation}
s
=
\frac{1}{Q(N_{+}+N_{-})}
\sum_{i=1}^{Q}
\sum_{j=1}^{N_{+}+N_{-}}
\left\|
\mathbf x_i-\mathbf y_j
\right\|_2,
\label{eq:feature_scale}
\end{equation}
where $\mathbf y_j\in\mathcal Y^{+}\cup\mathcal Y^{-}$ and
$\|\cdot\|_2$ denotes the Euclidean norm.
The post-training objective is then
\begin{equation}
\mathcal L_{\mathrm{Drift}}
=
\frac{1}{Q}
\sum_{i=1}^{Q}
\left\|
\frac{\mathbf x_i-\mathbf t_i}
{\operatorname{sg}(s)}
\right\|_2^2.
\label{eq:drift_loss}
\end{equation}
The target and normalization factor are detached during differentiation,
while gradients from $\mathbf x_i$ are backpropagated via the frozen
audio encoder and decoding pipeline to update only the generator parameters.

\section{Experiments}
\label{sec:experiments}

\subsection{Experimental Setup}
\label{sec:exp_setup}

\textbf{Dataset:}
We conducted all experiments on AudioCaps~\cite{kim2019audiocaps}.
Because some source audio was unavailable, our locally recovered subset contained $45{,}173$ training clips ($\sim125.5$ hours),
$445$ validation clips, and $877$ test clips.
The validation set was used for hyperparameter tuning and checkpoint selection, while the test set was evaluated only after the configuration was fixed.
All clips are processed as 10-s segments following MeanAudio~\cite{li2026meanaudio}.

\textbf{Implementation details:}
We applied the proposed DriftAudio framework to the MeanAudio-S-Full~\cite{li2026meanaudio} (denoted MeanAudio below) and FdAudio~\cite{huang2026fdaudio} checkpoints.
We used a frozen PANNs Cnn14 encoder~\cite{kong2020panns} at 16\,kHz to extract $2048$-dimensional audio features.
We precomputed PANNs features for all $45{,}173$ real training samples, from which $N_{+}=2048$ positive references are sampled without replacement at each optimization step.
We maintained $N_{-}=1024$ detached generated references in a FIFO bank initialized from the corresponding pretrained generator.
Each step uses $Q=64$ current-generated samples, whose detached pre-update features are used to refresh the FIFO bank.
We use AdamW~\cite{loshchilov2018decoupled} with a learning rate of $10^{-6}$, a weight decay of $10^{-6}$, and gradient clipping at $1.0$.
Checkpoint selection was based on validation FAD, yielding checkpoints after $4{,}400$ and $1{,}800$ optimization steps when starting from MeanAudio and FdAudio, respectively.
For memory efficiency, we computed the detached field without retaining the generator computation graph and exactly replayed the same text--noise pairs during backpropagation.

\textbf{Metrics:}
Following MeanAudio~\cite{li2026meanaudio}, we reported FAD in the VGGish feature space~\cite{kilgour2019fad,hershey2017cnn},
FD in the $2048$-dimensional PANNs feature space~\cite{kong2020panns},
KL and IS from PANNs outputs~\cite{salimans2016improved}, and CLAP as the mean text--audio cosine similarity from LAION-CLAP~\cite{wu2023clap}.
The lower the FAD, FD, and KL values, the better, while the higher IS and CLAP values, the better. 
We additionally reported latency for generating a batch of $8$
audio clips~\cite{huang2025impact} on an NVIDIA L40 GPU.

\begin{table*}[t]
    \centering
    \caption{
    Results of one-step TTA generation on AudioCaps.
    The upper block compares representative one-step TTA systems,
    while the lower block compares distributional post-training methods.
    }
    \label{tab:main_results}
    \begin{tabular*}{\textwidth}{
        @{\extracolsep{\fill}}lccccccc}
        \hline
        Method
        & NFE
        & FAD $\downarrow$
        & FD $\downarrow$
        & KL $\downarrow$
        & IS $\uparrow$
        & CLAP $\uparrow$
        & Latency (s) $\downarrow$ \\
        \hline

        \multicolumn{8}{l}{
        \textit{General one-step TTA systems}} \\

        ConsistencyTTA~\cite{bai2024consistencytta}
        & 1 & 2.25 & 22.58 & 1.39 & 8.91 & 0.287 & \textbf{0.875} \\

        SoundCTM~\cite{saito2025soundctm}
        & 1 & 1.94 & 18.68 & 1.31 & 8.58 & 0.275 & 0.964 \\

        MeanAudio~\cite{li2026meanaudio}
        & 1 & 1.68 & 15.43 & 1.27 & 10.36 & 0.322 & 1.271 \\

        \hline
        \multicolumn{8}{l}{
        \textit{Distributional post-training}} \\
        \textbf{DriftAudio (from MeanAudio)}
 & 1 & 1.11 & 12.71 &  \textbf{1.21} & 9.83 & 0.325 & 1.271\\

        FdAudio~\cite{huang2026fdaudio}
& 1 & 1.22 & 13.73 & 1.27 & \textbf{10.84}
& \textbf{0.352} & 1.271 \\

\textbf{DriftAudio (from FdAudio)}
& 1 & \textbf{0.99} & \textbf{12.61} &  1.24 & 10.17 & 0.335 & 1.271 \\

        \hline
    \end{tabular*}
\end{table*}

\begin{figure}[t]
    \centering
    \includegraphics[width=\columnwidth]{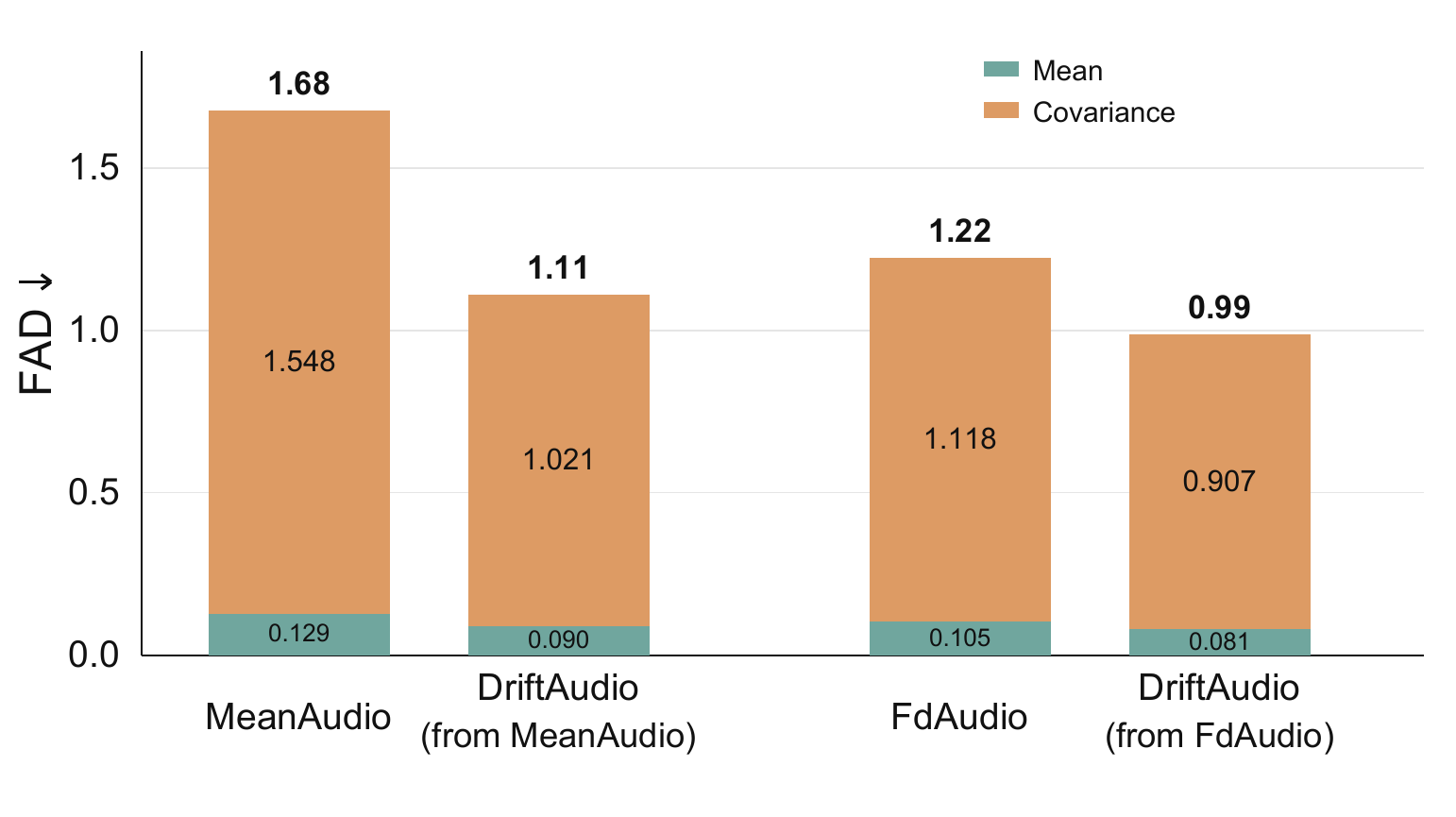}
    \caption{
Mean and covariance components of FAD for the four evaluated models.
Numbers above the bars denote total FAD.
    }
    \label{fig:fad_decomposition}
\end{figure}

\subsection{Main Results and Comparisons}
\label{sec}

We compared DriftAudio with representative one-step TTA systems and an existing distributional post-training method.
All comparison models were evaluated from their publicly released checkpoints, including ConsistencyTTA~\cite{bai2024consistencytta}, SoundCTM~\cite{saito2025soundctm}, MeanAudio~\cite{li2026meanaudio}, and FdAudio~\cite{huang2026fdaudio}.
In particular, FdAudio is obtained by further post-training MeanAudio on an unreleased $80{,}000$-sample subset randomly drawn from the AudioCaps--WavCaps pool~\cite{huang2026fdaudio,mei2024wavcaps}.
All baseline models were reevaluated on the same locally recovered test set for consistency, and may therefore differ modestly from those reported in the original publications.

Table~\ref{tab:main_results} summarizes the results.
The upper block compares representative one-step TTA systems, while the lower block focuses on distributional post-training.
All results are reported with NFE $=1$.
Since FdAudio and DriftAudio modify only the model weights during post-training, they preserve the inference procedure of MeanAudio and introduce no additional inference-time computation.

DriftAudio (from MeanAudio) and DriftAudio (from FdAudio) denote post-training starting from the corresponding checkpoints.
Overall, DriftAudio consistently improves FAD, FD, and KL from both starting checkpoints.
Starting from MeanAudio, it reduces FAD from $1.68$ to $1.11$ and FD from $15.43$ to $12.71$, while achieving the lowest KL of $1.21$.
Starting from FdAudio, it further reduces FAD from $1.22$ to $0.99$ and FD from $13.73$ to $12.61$.
These gains are accompanied by trade-offs in IS and CLAP: when starting from MeanAudio, CLAP improves slightly while IS decreases; when starting from FdAudio, both IS and CLAP decrease.
These results demonstrate the effectiveness of DriftAudio as a distributional post-training method for one-step TTA generation.

DriftAudio performs in the PANNs feature space, whereas FAD is computed using VGGish features.
We further analyze the mean and covariance components of FAD to examine whether the distributional improvement extends beyond the feature space used for post-training.
Fig.~\ref{fig:fad_decomposition} shows that DriftAudio reduces both components from both starting checkpoints.
Starting from MeanAudio, the covariance decreases substantially from $1.548$ to $1.021$, together with a reduction in the
mean from $0.129$ to $0.090$.
A similar trend is observed when starting from FdAudio.
These results show that the distributional improvement is also observed in the VGGish feature space, which is not used by the DriftAudio objective.

\subsection{Analysis and Ablation Study}
\label{sec:ablation}

\begin{table}[t]
    \centering
    \setlength{\belowcaptionskip}{4pt}
    \caption{
Text--audio alignment on generated samples. 
    }
    \label{tab:conditional_preservation}
    \begin{tabular*}{\columnwidth}{
        @{\extracolsep{\fill}}lccc}
        \hline
        Method
        & CLAP $\uparrow$
        & R@1 $\uparrow$
        & R@5 $\uparrow$ \\
        \hline
        MeanAudio
        & 0.322 & 32.5 & 71.0 \\
        DriftAudio (from MeanAudio)
        & 0.325 & 34.5 & 74.2 \\
        FdAudio
        & \textbf{0.352} & \textbf{38.7} & \textbf{74.9} \\
        DriftAudio (from FdAudio)
        & 0.335 & 36.7 & 73.1 \\
        \hline
    \end{tabular*}
\end{table}

\textbf{Text--audio alignment.}
Since DriftAudio estimates a marginal drifting field across text conditions, we examined whether marginal post-training preserves the dependence of generated audio on its conditioning text.
Following audio--text retrieval evaluation~\cite{Oncescu21a}, we treated each generated audio sample as a query and ranked candidate captions by cosine similarity in the CLAP embedding space.
Duplicate captions in the test set were merged, resulting in $866$ unique candidate captions.
We reported audio-to-text Recall@1 (R@1) and Recall@5 (R@5), indicating whether the matched caption appears among the top $1$ or top $5$ retrieved candidates

Table~\ref{tab:conditional_preservation} shows that FdAudio achieves the highest scores across all three metrics, including an R@1 of $38.7\%$.
Starting from MeanAudio, DriftAudio improves both CLAP and retrieval performance, increasing R@1 from $32.5\%$ to $34.5\%$.
Starting from FdAudio, DriftAudio shows a modest decrease relative to FdAudio but remains above MeanAudio on all three metrics.
Overall, the retrieval results suggest that marginal distributional post-training does not substantially weaken text--audio alignment.

\textbf{Ablation study.}
We evaluated a continued-training baseline and three DriftAudio design
choices, all starting from the released FdAudio checkpoint and using
the same validation-based checkpoint selection protocol.
The \textbf{default configuration} used per-step real-reference
resampling, PANNs features, $N_{+}=2048$ real references, and
$N_{-}=1024$ generated references.
We compared it with:
(1) \textbf{Continued FdAudio training}, which continued post-training
using the original FdAudio objective;
(2) \textbf{Fixed real references}, which used the same
$N_{+}=2048$ real references throughout training;
(3) \textbf{AudioMAE features}, which replaced PANNs with
AudioMAE~\cite{huang2022audiomae}; and
(4) \textbf{Larger reference sets}, which doubled the reference sizes
to $N_{+}=4096$ and $N_{-}=2048$.
For each variant, we selected the checkpoint with the lowest validation
FAD.

Table~\ref{tab:method_ablation} shows that the default configuration provides a practical and effective setting: real-reference resampling improves FAD and FD, the drifting formulation remains effective in a different audio feature space, and increasing the reference-set size brings only marginal gains.
Continued FdAudio training slightly improves FD and CLAP but worsens FAD from $1.22$ to $1.40$, whereas DriftAudio reduces it to $0.99$.
This control shows that the gains cannot be explained by additional optimization alone, supporting the marginal Drifting objective.
Using fixed real references increases FAD from $0.99$ to $1.04$.
Replacing PANNs with AudioMAE still yields competitive performance, suggesting that the formulation can operate across different audio feature spaces, although PANNs performs better in the current setting.
Doubling the reference-set sizes yields a comparable FAD of $0.98$, but increases the median training time per optimizer step from $23.7$ s to $49.5$ s on a single NVIDIA L40 GPU ($2.09\times$).
The default reference sizes provide a practical balance between distributional performance and training cost.
% Overall, these results support both the marginal Drifting objective and the adopted DriftAudio configuration.

\begin{table}[t]
\centering
\caption{
Ablation of key DriftAudio design choices and continued-training baseline from FdAudio.
}
\label{tab:method_ablation}
\begin{tabular*}{\columnwidth}{
@{\extracolsep{\fill}}lccc}
\hline
Variant
& FAD $\downarrow$
& FD $\downarrow$
& CLAP $\uparrow$ \\
\hline
Continued FdAudio training
& 1.40 & 13.57 & \textbf{0.358} \\
\hline
Fixed real references
& 1.04 & 12.94 & 0.343 \\

AudioMAE features
& 1.02 & 13.41 & 0.333 \\

Larger reference sets
& \textbf{0.98} & 12.69 & 0.338 \\

\textbf{Default configuration}
& 0.99 & \textbf{12.61} & 0.335 \\
\hline
\end{tabular*}
\end{table}

\section{Conclusion}
\label{sec:conclusion}

We introduced DriftAudio, a distributional post-training method for pretrained one-step TTA generators that applies Drifting to the marginal audio-feature distribution.
DriftAudio retains text-conditioned generation while estimating the drifting field from real and generated samples collected across text conditions.
On AudioCaps, DriftAudio consistently improves FAD, FD, and KL when starting from both MeanAudio and FdAudio, without substantially weakening text--audio alignment.
These results suggest that marginal Drifting may provide a general distributional post-training principle for conditional generative models.
Future work will extend the current single-encoder formulation to multiple complementary audio representations and include subjective perceptual evaluation of the generated audio.

\clearpage
% References should be produced using the bibtex program from suitable
% BiBTeX files (here: strings, refs, manuals). The IEEEbib.bst bibliography
% style file from IEEE produces unsorted bibliography list.
% -------------------------------------------------------------------------
{\footnotesize
\bibliographystyle{IEEEbib}
\bibliography{refs}
}

\end{document}